Atmospheric $CO_2$ and total electricity production before and during the nation-wide restriction of activities as a consequence of the COVID-19 pandemic

[1,3]Yusup Y., [2*]Kayode, J.S., [1]Mardiana Idayu Ahmad, [3]Chee Su Yin, [4]Muhammad Sabiq Mohamad Nor Hisham, [4]Hassim Mohamad Isa

[1]Environmental Technology, School of Industrial Technology, Universiti Sains Malaysia, 11800 USM, Pulau Pinang, Malaysia.

[2]Department of Research and Innovations, Institute of Hydrocarbon Recovery, Universiti Teknologi PETRONAS, Persiaran UTP, 32610 Bandar Seri Iskandar, Perak Darul Ridzuan, Malaysia.

[3]Centre for Marine and Coastal Studies (CEMACS), Universiti Sains Malaysia, Pulau Pinang, Malaysia.

[4]A-LG-03, Block A, Serdang Perdana Selatan, Section 1, 43300 Seri Kembangan, Selangor Darul Ehsan, Malaysia.

*Corresponding Author: jskayode@gmail.com, john.kayode@utp.edu.my

**Abstract**

In this paper, we analysed real-time measurements of atmospheric $CO_2$ with total electricity production and nation-wide restrictions phases due to the novel coronavirus COVID-19 pandemic, and its effects on atmospheric $CO_2$ concentrations. A decline of 3.7% in the global energy demand at about 150 million tonnes of oil equivalent (Mtoe) in the first quarter (Q1), of 2020 was recorded, as compared to the same first quarter (Q1), of 2019, due to the cutback on global economic activities. Our results showed that: 1) electricity production for the same period in the years 2018, 2019, and 2020, shrunk at an offset of about 9.20%, which resulted in the modest reduction of about (-1.79%), in the atmospheric $CO_2$, to that of 2017-2018 $CO_2$ level; 2) a non-seasonal abrupt; but brief, atmospheric $CO_2$ decrease by about 0.85% in mid-February 2020, could be due to the Phase 1 movement restrictions in China. The results showed that, the reduction in electricity production is significant to the short-term variability of atmospheric $CO_2$. It also highlights the significant contributions from China to the atmospheric $CO_2$, which suggests that, without the national restriction of activities, $CO_2$ concentration are set to exceed 2019 by 1.79%, but it quickly decreased due to the lockdown, and sustained the reduction for two consecutive months. The results underscore the atmospheric $CO_2$ reductions on the monthly time scale that can be achieved, if electricity production from combustible sources were slashed, which could be useful for cost-benefit analyses of the reduction in electricity production from combustible sources, and the impact of these reduction to the atmospheric $CO_2$.

1. **Introduction**

1.1 COVID-19 Restrictions and its Environmental Impact

The coronavirus derived its name from the crown of sugary-proteins that envelope the corona atom (Ather, et al., 2020; Chang, et al., 2020; Han, et al., 2020). The symptoms of this disease is similar to influenza or the common cold with the accompanying cough, fever and breathing difficulties. The disease was not globally-spread until December 2019 when it was detected in the city of Wuhan, China. After the Wuhan outbreak, the disease spread exponentially across all nations of the world claiming hundreds of thousands of lives, especially among the elderly and immunocompromised individuals (Ather, et al., 2020; WHO, 2020). In Malaysia, the COVID-19 index case was detected on 25 January 2020. Within a few weeks, the number of detected cases rose sharply (Abdullah et al., 2020), particularly in March and extended to the end of April 2020.

The COVID-19 pandemic has led to a public health predicament of unprecedented magnitude, and aside from research work on developing treatment and vaccine, some studies have gone to address the effects of the "lockdown" due to the pandemic on the environment, (i.e., Abdullah et al., 2020, IEA, 2020, Le Quéré, et al., 2020, Myllyvirta, 2020). The large scale on which the phenomenon has occurred offers a rare glimpse on the effect of policies, and strategies, on the global atmosphere that would otherwise be difficult to study. One such global policy is nationwide restrictions on industrial; educational, and tourism activities, that resulted in lower energy demand, and reduced greenhouse gas emissions, such as the $CO_2$ (CarbonBrief, 2020; Le Quéré et al., 2020). This abrupt decline in emissions, could lead to decreased in the atmospheric $CO_2$ concentrations of which this work intends to explore.

In the face of the pandemic, most governments around the world enforced directives, and legislate orders restricting mobility, gatherings, or meetings. Workers all over the world were ordered to work from home due to the confinement imposed aftermath of the COVID-19 pandemic outbreak. The restrictions in movement across nations, states and cities, negatively impacted manufacturing, agriculture, shipping, and tourism sectors. The confinement of people to their homes, and with all the industries and factories shuttered, it was observed that drastic reduction from vehicular, and industrial emissions had a positive impact on the air quality, e.g., (Abdullah et al. 2020), and could extend to an appreciable reduction in the atmospheric $CO_2$ concentrations.

A report released on the CarbonBrief website, (www.carbonbrief.org) on 19 May 2020, showed a global reduction of the daily $CO_2$ emissions at 17%, below that of the 2006 across 69 countries, (CarbonBrief, 2020; Le Quéré, et al., 2020; Peters, et al., 2020). In the recently published work by Abdullah et al. (2020), on the air quality status during the Malaysia's Movement Control Order (MCO), the particulate matter's concentrations showed up to 58.4% reduction in concentration, in several of the areas marked as COVID-19 pandemic "red zones". A similar pattern was observed in the emission of $CO_2$, i.e., (CarbonBrief, 2020; Le Quéré et al., 2020). Although, the effect of reduced $CO_2$ emissions from the atmospheric $CO_2$ has yet to be reported.

In this paper, we studied whether global movement restrictions, and complete shutdown of most economic sectors, and subsequent decrease in electricity production, and $CO_2$ emissions, reduced the atmospheric $CO_2$ concentrations. Atmospheric $CO_2$ concentration was collected from a station situated in a remote location, and the data was then compared with the different stages of the restriction of activities imposed by high-emitting $CO_2$ nations of the world, like the China, Europe, the United States of America, and India, and was evaluated against the electricity production data, to determine if the latter affected atmospheric $CO_2$ concentrations, or not.

## 2. Data and Methods

Real-time atmospheric $CO_2$ concentration data was measured from a weather station on a tropical coast where other parameters such as the mass and energy fluxes, micro-climate variables, sea surface and water temperature data are continuously collected, and computed. The station is located at the Centre for Marine and Costal Studies (CEMACS), Universiti Sains Malaysia, (USM), at the north-western end of the Pulau Pinang Island, in Peninsula Malaysia. This station can be considered, as a station that measures background $CO_2$, because of its minimally influenced by the $CO_2$ from anthropogenic sources. The research station is situated on latitudes 5° 28' 06" N, and longitudes 100° 12' 01" E. The station was named the ''Muka Head Station'', where the instruments were mounted on a stainless-steel podium that was built in 2015, and expanded on a pre-existing jetty. Further details on the station can be found in the published literature, e.g., (Yusup et al., 2018a,b,c, 2019), and the station website at, http://atmosfera.usm.my. The station is in the distinctive intertidal zone of a coastal ocean, on the tropical continental shelf along the Straits of Malacca, in the southern South China Sea, and can be classified as a narrow and shallow continental shelf (Cai et al., 2006).

Atmospheric $CO_2$ concentration in parts per million (ppm) was measured using an infrared gas analyser, (model LI-7550, LI-COR, USA), at a frequency of 20 Hz. The high-frequency data was quality-checked, and processed, using the EddyPro© software, (version 6.2.0, LI-COR, USA), and the Tovi© software, (version 2.8.1, LI-COR Biosciences, USA). Further analyses were performed, and plots were produced using the open source R statistical software, (version 3.6.3), and the RStudio, (version 1.2.5033).

The different phases of the world-wide restriction of activities, is defined in Table 1, which corresponds to the decreased in the industrial activities, and electricity production, which have been associated with the $CO_2$ emissions. Here, only the electricity produced by coal, natural gas, and other combustibles are considered. This classification was adapted and simplified, from the report of Le Quéré et al. (2020), where restricted activities were said to have the potential of reducing the $CO_2$ emissions, and thus, were categorized.

The global energy demands for the first quarter in 2020, (Q1 2020), was analysed using the available data from the Global Energy Review, (GER), for the year 2020, e.g., (IEA, 2020). The results showed energy demand reduction in some selected countries, after the implementation of the stricted lockdown. Total electricity production for the months of November 2019 to May 2020, was obtained alongside the energy demands data from the GER.

Phase 1 is for the China lockdown, which has the largest drop in the electricity production at -12% in February 2020, as compared to the February 2019, e.g., (IEA, 2020), in addition to China's contributions of 28% in the global total $CO_2$ emissions, (IEA, 2019). China was the first country to implement the national lockdown, at the end of January 2020, which led to the nationwide restrictions of activities. By early March 2020, other regions around the world, also charted positive, and negative changes; e.g., the United States, with +0.002% electricity production at 14% of global $CO_2$ emissions; Europe, (Germany, United Kingdom, Italy, France, and Poland, with -13% electricity production at 6% global $CO_2$ emissions), India with +16% electricity production at 7% of global $CO_2$ emissions, and other countries, where their respective governments enforced national restriction on all activities, which further decreased electricity production, and $CO_2$ emissions. This duration is categorized as Phase 2. In the local front, the Malaysian government enforced the Movement Control Order (MCO), on 18 March 2020, that obligated most industries, and factories, except essential services like the healthcare, and food industries, to shut down, and workers were ordered to work from home.

Table 1. Definitions used for the different phases of the world-wide restriction of activities.

| Phase | Countries Imposing Nation-wide or *Multiple-States Restrictions. |
|---|---|
| Phase 1 | National restriction of activities for China (-12% electricity production, and 14% global $CO_2$ emissions). |
| Phase 2 | National restrictions of activities for *the United States, Europe (Germany, United Kingdom, Italy, France, and Poland), and India (+1% electricity production, and 27% global $CO_2$ emissions). |
| MCO | Movement Control Order, the national restriction of activities for Malaysia (-50% electricity production, and >1% global $CO_2$ emissions). |

## 3. Results and Discussion

*3.1 Reduction in global energy demand and electricity production during COVID-19 pandemic.*

Energy is the input of all economic activities, and the expenditure of energy releases the $CO_2$ to the atmosphere. The COVID-19 pandemic, is anticipated to lower the global energy demands, and consumption, and thus decreases the $CO_2$ emissions. A decline of 3.7%, at about 150 million tonnes of oil equivalent, (Mtoe), in the first quarter (Q1), of 2020 global energy demand, was reported as compared to the Q1 of 2019, due to the cutback of global economic activities in various sectors, i.e., (IEA, 2020). The percentage of declined by sources, and the contributed factors, is summarized in Table 2.

Table 2. Percentage of global energy decline by source, and the contributed factors in Q1 of 2020 compare to the Q1 of 2019, (IEA, 2020).

| Energy demand by source | Q1 2020 change from Q1 2019 (%) | Contributed factors |
| --- | --- | --- |
| Coal | -8 | Reductions in electricity demand |
| oil | -5 | Restrictions on travel, reduced demand of vehicles |
| Natural gas | -2 | China and Europe experiencing the most significant declines, due to low consumer demands. |

Energy demand has declined significantly because of full strict, and partial lockdowns, imposed by governments around the world. The International Energy Agency, (IEA), estimated sinks in energy demands of -15% in countries where comprehensive, and strict lockdowns, have been enforced, (IEA, 2020). Currently, up-to-date monthly energy demands data is unavailable to be directly compared with the atmospheric $CO_2$ concentrations, as a result, the total electricity production, (TEP), is used as proxy to the energy demands.

The global, or total demands of energy in the form of electricity produced, from coal, natural gas, and other combustibles sources, for the months of November to May for the three years periods, i.e., 2017-2018, 2018-2019, and 2019-2020 is presented in Fig. 1. Phase 1 covers the period for the national restrictions of activities in China, while the Phase 2 covers periods of lockdown for the other parts of the world, i.e., the United States, Europe, and India. The data shows the seasonal decline in the TEP during the months of January until May, in each of the periods considered (i.e., from the peak production of about 580 TWh, at the end of December, to 470 and 490 TWh, in the 2017-2018, and 2018-2019, respectively). The 2019-2020 TEP adhere to the same trend, although for the same period, it is consistently lower by 100 +/- 50 TWh, from 485 TWh to 420 TWh. At the time of writing this paper, no data was available for the months of February to May 2020.

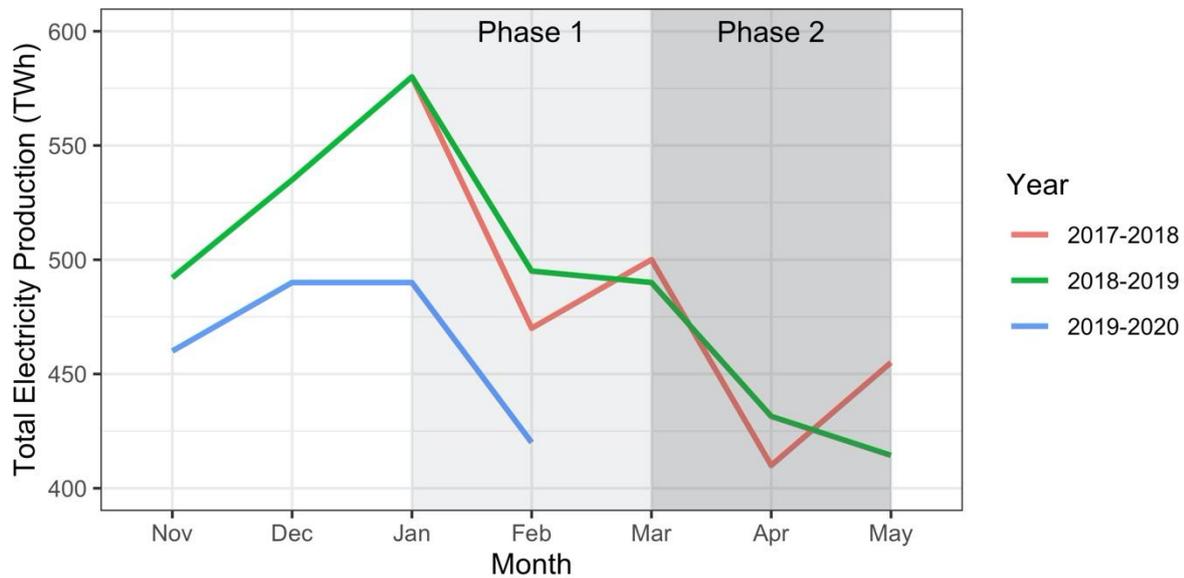

Fig. 1. Total electricity production, (TEP), for the months of November to May for three years periods; Phase 1 is the duration for the national restrictions of activities in China, while Phase 2 is for parts of the United States, Europe, and India.

From the limited data, it shows that Phase 1 and Phase 2 did not affect the trend of TEP, but offsets it, to a lower average, at approximately 9.20% from the TEP, if restrictions or lockdowns were not enforced. Referring to Table 1, the offset is also triggered by the lockdown of China (Phase 1), while Phase 2 did not further slash TEP. The 9.20% decline is consistent with the overall decrease in energy demand reported by the International Energy Agency (IEA, 2020). The MCO is unlikely to significantly decrease the TEP since Malaysia is a small country with low TEP of about 12000 to 13000 GWh monthly (IEA, 2020). Other reports stated that the significant decrease of over 5% in the Q1 of 2020 global energy demand has brought about a substantial decline in the $CO_2$ emissions as compared to the Q1 of 2019, which stem from disruptions of transport systems and industrial sectors of the economy, as well as reduction in products and services demands, (IEA, 2020, Le Quéré et al., 2020, Myllyvirta, 2020). In Le Quéré et al. (2020), an even more rapid decrease was projected across the remaining nine months of 2020 with the predicted reduction value of about 30.6 Gt of carbon. However, as shown in Fig. 1, our results show that the cutback in TEP might unfold as an offset reduction, and not a compounded drop in the TEP.

*3.2 Atmospheric $CO_2$ concentrations trends during COVID-19*

The monthly-averaged of atmospheric $CO_2$ concentrations in 2019-2020, was lower than that of 2017-2018, and 2018-2019. The differences in the $CO_2$, using the 2019-2020 as base period, is presented in Table 2. The $CO_2$ recorded an average of 408.69 ppm for the month of January 2020, (i.e., Phase 1), with a maximum difference of -2.36 ppm, e.g., (-1.67%), for the 2018-2020, and 11.85 ppm, e.g., (+0.68%), for the 2019-2020. In March, (i.e., Phase 2), the world witnessed strict lockdowns in most places around the world, particularly, the United States of America, Europe, and India. There was reductions in the $CO_2$ ppm recorded, with -7.83, -24.52, and -3.32; for the mean, minimum, and maximum values, respectively, for the 2018-2020, which is a decline of -1.91%, from the 2018. Yet, the $CO_2$ ppm recorded for the 2019-2020 period was, -1.91, -17.88, and 6.54, for the respective mean, minimum, and maximum values, which is a decrease of 0.47% from 2019. The resuming of activities in March 2020, due to the reduction of movement restrictions in China, suggests that it may have an influence on the increasing atmospheric $CO_2$ in April. However, this trend is consistent with previous periods, which indicated that, the increase is a yearly periodic feature, and not due to increased activities after the easement.

TEP trends, reflect the changes in monthly-averaged atmospheric $CO_2$, e.g., (Fig. 1, and Table 2), but the largest negative differences, only occurred during the period 2018-2020 at, -1.67%, and -1.91%, (i.e., an average of 1.79%), in January, and March, respectively. This was not mirrored by the differences in the 2019-2020, where it was lower in December, and March, but increased by approximately 0.10%, during the same months of the lockdown. This growth suggests that the 2019-2020 restrictions, had only a modest impact on the atmospheric $CO_2$. One reason could be due to a cut in $CO_2$ emissions during the year 2019, due to the widespread adoption of renewable energy technologies in Europe, (IEA, 2020). The year 2019, was also the first year when the $CO_2$ emissions did not records increase at the rate of 2% per annum, (Le Quéré et al., 2020). Furthermore, atmospheric $CO_2$ has an average residence time of ten years, in the atmosphere, and so an abrupt decrease in anthropogenic emissions might not substantially decrease overall $CO_2$ concentrations, (Ballantyne et al., 2012). Therefore, decrease in the $CO_2$ level for the 2019-2020, (i.e., 413.55 ppm), is an indication that the lockdown only resulted in modest reduction of the atmospheric $CO_2$, slightly lower when compared to the 2017-2018 $CO_2$ level, (i.e., 414.32 ppm), and a little lower than the 2018-2019, (i.e., 413.88 ppm) $CO_2$ level.

Table 2. Monthly-averaged differences in the atmospheric $CO_2$ concentrations (ppm), for the 2019-2020, and 2018-2020; using the year 2020 as the base year.

| Month | 2018-2020 | | | | 2019-2020 | | | |
|---|---|---|---|---|---|---|---|---|
| | Mean | Min | Max | % | Mean | Min | Max | % |
| Nov | 1.09 | -17.83 | 5.89 | 0.26 | -1.16 | -22.89 | 4.57 | -0.28 |
| Dec | 0.06 | -15.87 | 4.55 | 0.01 | -7.41 | -20.33 | -3.79 | -1.81 |
| Jan | -6.84 | -20.94 | -2.36 | -1.67 | 2.79 | -16.75 | 11.85 | 0.68 |
| Feb | 4.40 | -13.90 | 11.78 | 1.07 | 3.64 | -10.66 | 13.30 | 0.88 |
| Mar | -7.83 | -24.52 | -3.32 | -1.91 | -1.91 | -17.88 | 6.54 | -0.47 |
| Apr | 3.43 | -6.93 | 7.29 | 0.81 | 0.66 | -13.75 | 4.26 | 0.16 |
| May | 8.99 | 6.03 | 8.98 | 2.10 | 10.22 | 5.49 | 14.72 | 2.38 |

To see if there are short-term changes in concentration due to the lockdown, the daily-averaged of the $CO_2$, for Phase 1, Phase 2, and the MCO was analysed, (i.e., Fig. 2). The $CO_2$ concentration recorded, was slightly above 408.92 ppm in the month of March, then rose to 424.15 ppm, and 428.59 in the months of April, and May 2020, respectively. Before and during Phase 1, there were consistent downward trends during the middle of January for the 2017-2018, 2018-2019, and 2019-2020, that aligns with yearly recurring trends. The concentration then increased after February, which are also due to the periodicity of yearly $CO_2$ trends. The effect of the offset in TEP, for the 2019-2020, is not apparent here, e.g., (Fig. 1). Although the lockdown in Malaysia, was comprehensively severe, and strict, with dropped in TEP to about 50%, from February 2020, to April 2020. The Malaysian MCO showed little effect on the $CO_2$ concentration as expected. Hence, the $CO_2$ trends reported here are large-scale trends, and not sensitive to local, or even regional fluctuations.

Aside from electricity production, meteorological parameters may explain some of the seasonal patterns of the $CO_2$ concentration. The negative association between atmospheric $CO_2$, and the atmospheric temperature in the first quarter of the year, is due to the effects of radiation perturbation caused by induction of aerosol, e.g., (Lee, et al., 2020; Xie et al., 2020; Zhang, et al., 2020). The $CO_2$ trends here, supported the results from these authors with the reduction of $CO_2$ concentration recorded for the period of dryness occasioned by excessive solar radiation, and the absence, or low precipitation along the western coastline of Peninsula Malaysia. Intense precipitation, has been characterizing the local atmosphere since the middle of March, occasioned by the Spring Transitional Monsoon, (STM), that correlates with drops in atmospheric temperatures, which subsequently raised the atmospheric $CO_2$. Previous reports by Yusup et al. (2019), gave the same positive correlation during the same period.

A feature not evident in the monthly-averaged data, is the abrupt but brief $CO_2$ concentration decrease, in the mid-February 2020, e.g., (Fig. 2). This 0.85% dip, is not apparent in the previous periods of the years 2017-2018, and 2019-2020, and do not correlate with any atmospheric temperature. The dip could be due to the Phase 1, (i.e., China's massive lockdown), in which reduction of TEP in 2020 was 12% lower than the year 2019. The following low $CO_2$ concentration, in the mid-February, and March 2020, could be due to the Phase 2. Although, not reported here, the lockdown have also impacted other anthropogenic emissions primarily sourced from the combustion of fossil fuels, such as the tourism and shipping subsectors of the economy, which further slashed the $CO_2$ emissions on top of the reduction in the TEP. After two months, the $CO_2$ concentration approached, and surpassed the season-averaged concentration in April 2020. Hence, contributions from the 12% decline in the TEP in China has a short-term influence, on the reduction of atmospheric $CO_2$.

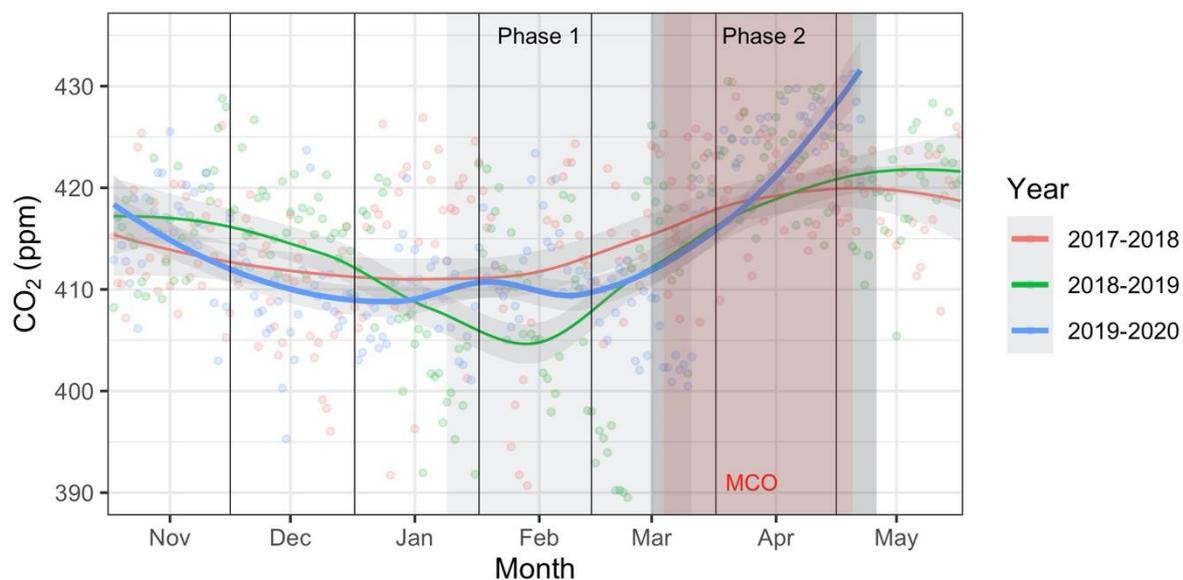

Fig. 2.   Atmospheric $CO_2$ concentration for the months of November to May for three years; Phase 1 is the duration of national restrictions of activities in China; while Phase 2, is for other parts of the United States, Europe, and India; the Movement Control Order (MCO), is the national restrictions rule in Malaysia.

The difference in $CO_2$ concentration for the years 2018-2019, and 2019-2020, emphasized that the reduction caused by the national restrictions of activities, is for both the years 2018-2020, and 2019-2020, (i.e., Fig. 3), where it occurred in two instances, e.g., December-January, and March. The large decline in December 2019, for the years 2019-2020, is swiftly compensated in the January 2020, but the peak was suppressed by the Phase 1 restrictions in China. After the Phase 1 ended, the $CO_2$ differences began to rise steadily, while the restrictions in the Phase 2 did not seems to have affected the latter.

Again, the Malaysia's MCO did not affect the $CO_2$ differences. This is largely due to the overall TEP in China, that declined by 12%, while for the Phase 2 countries, it increased by 1%. These results emphasizes the significant contributions of China to the atmospheric $CO_2$ concentration. The results also suggest that, without the national restrictions of activities, the $CO_2$ concentrations are set to exceed the year 2019 values by 1.79%, but due to the lockdown, it quickly decreased, and sustained the lower values for two consecutive months before climbing again in the mid-April 2020.

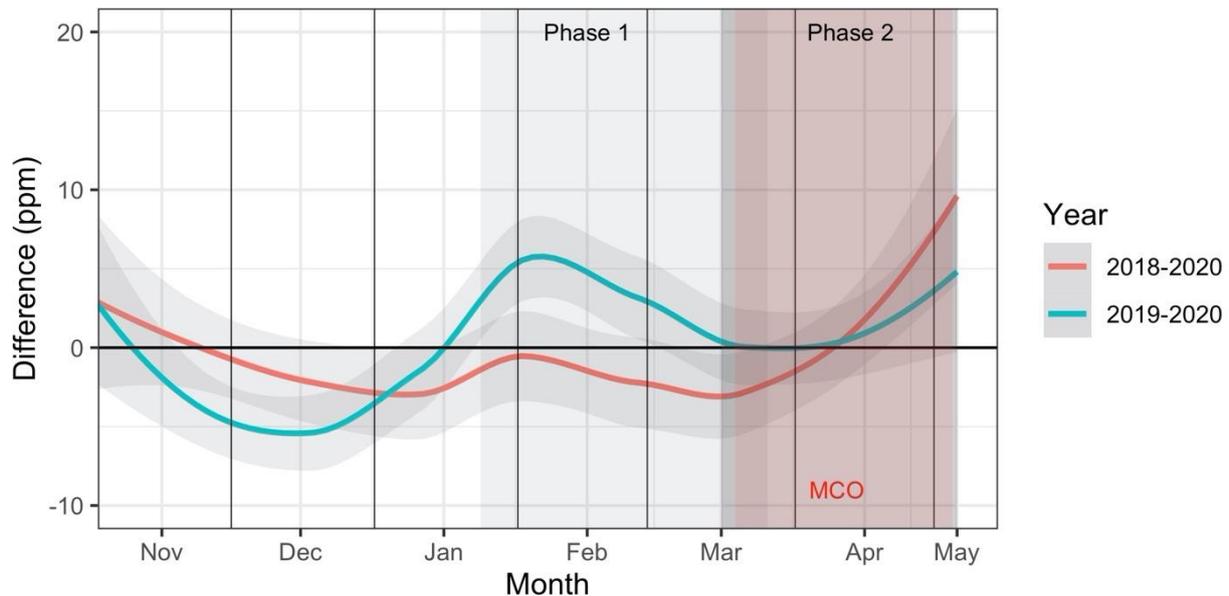

Fig. 3. Differences in the atmospheric CO2 concentrations for the years 2018-2020, and 2019-2020 periods for the months of November to May.

## 4. Conclusions

Our analysis reports on effects of the nationwide restrictions of activities, decreased in the electricity generations from the combustible sources, on monthly-, and daily-averaged atmospheric $CO_2$ concentrations. The results shows that the electricity production for the same period in the years 2018, 2019, and 2020, shrunk at an offset of about 9.20%. This resulted in the modest reduction, e.g., (-1.79%) of the atmospheric $CO_2$, and in the December-January, and March, from 414.32 ppm in 2017-2018 periods. It only slightly decreased from the 2018-2019 $CO_2$ concentrations level, e.g., (413.88 ppm), likely due to the breakthroughs in the renewable energy technologies, and the increased in adoption of the renewable energy sources. A novel feature observed here, is the non-seasonal, and abrupt, but brief atmospheric $CO_2$ decrease of about 0.85%, in the mid-February 2020, due to the Phase 1 restrictions in China. Hence, the reduction in electricity production is significant on a short-term variability of the atmospheric $CO_2$ concentrations. It also highlights the significant contributions from China to the atmospheric $CO_2$, which suggests that, without the national restrictions of activities, $CO_2$ concentrations was projected to exceeded that of the year 2019 by 1.79%, but thanks to the severe lockdown, it quickly decreased to the lower level, and sustained the low values for two months successively. The results underscore the atmospheric $CO_2$ reductions on the monthly time-scale that can be achieved, if the electricity productions from combustible sources were slashed, which could be useful for cost-benefit analyses of the reduction in the electricity production from these combustible sources, and the impact of these reductions on the atmospheric $CO_2$ concentrations.

## Author Contribution Statement

**Yusup Yusri:** conceptualization, investigation, writing, software, formal analysis

**Kayode John Stephen:** conceptualization, original draft writing and editing final manuscript

**Mardiana Idayu Ahmad:** writing, methodology

**Chee Su Yin:** reviewing and editing

**Muhammad Sabiq Mohamad Nor Hisham:** resources

**Hassim Mohamad Isa:** resources

## Competing Interests

The authors declare that they have no known competing financial interests or personal relationships which have, or could be perceived to have, influenced the work reported in this article.

## Acknowledgments

The authors thank those who contributed to the success of this work. We specially acknowledge the efforts of the reviewers to better the readability of this paper.